\documentclass[vecphys]{svmult}

\usepackage{makeidx}         
\usepackage{graphicx}        
\usepackage{listings}
\usepackage{color}
\usepackage{multicol}        
\usepackage[bottom]{footmisc}

\makeindex             

\usepackage{amsmath}
\usepackage{amsfonts}

\begin{document}

\title*{Performance limitations for sparse matrix-vector multiplications 
   on current multicore environments}
\titlerunning{Performance limitations for sparse matrix-vector multiplications}
\author{Gerald Schubert \and Georg Hager \and Holger Fehske}
\authorrunning{G. Schubert et al.}
\institute{
G. Schubert $\cdot$ G. Hager
\at Regionales Rechenzentrum Erlangen, Friedrich-Alexander Universit\"at Erlangen-N\"urnberg, 
Martensstr. 1, D-91058 Erlangen, Germany\\
\email\texttt{\{gerald.schubert,georg.hager\}@rrze.uni-erlangen.de}
\and
H. Fehske
\at
Ernst-Moritz-Arndt-Universit\"at Greifswald,
Institut f\"ur Physik, Felix-Hausdorff-Str. 6, \\
D-17489 Greifswald, Germany\\
\email \texttt{fehske@physik.uni-greifswald.de}
}

\maketitle

\abstract{
The increasing importance of multicore processors calls for a reevaluation of
established numerical algorithms in view of their ability to profit from this
new hardware concept.
In order to optimize the existent algorithms, a detailed knowledge of the 
different performance-limiting factors is mandatory.
In this contribution we investigate sparse matrix-vector multiplication,
which is the dominant operation in many sparse eigenvalue solvers. 
Two conceptually different
storage schemes and computational kernels have been conceived in the past to target
cache-based and vector architectures, respectively.
Starting from a series of microbenchmarks we apply the gained insight
on optimized sparse MVM implementations, whose serial and OpenMP-parallel
performance we review on state-of-the-art multicore systems.  
}

\section{Introduction}

Large sparse eigensystems or systems of linear equations emerge 
from the mathematical description of many problems in physics,
chemistry, and continuum dynamics. Solving those systems often
requires multiplication of a sparse matrix with a vector as the
dominant operation. 
Depending on the matrix dimension, its sparsity and the complexity of the 
remaining algorithm, the fraction spent in the sparse matrix-vector 
multiplication (SpMVM) may easily constitute over 99\% of total run time.
An efficient 
implementation of the SpMVM on current hardware is thus fundamental,
and complements the development of sophisticated high-level algorithms 
like, e.g., Krylov-subspace techniques~\cite{Sa03}, preconditioners~\cite{De97,Gr97}, 
and multi-grid methods~\cite{Br81,BHM00}\@. 

The exponential growth in available computational power enables larger
problems, finer grids and more complex physics to be tackled on
present-day supercomputers. Distributed-memory parallelization has
always been instrumental in exploiting the full potential of sparse
solvers on massively parallel architectures, but with the advent of
multicore processors the per-core performance has begun to decline,
and parallel computers have developed a strongly hierarchical
structure.  Efficient implementations of the SpMVM must hence be
adapted to the new shared-memory and cache topology complexities as
well.  These optimizations will be the focus of this contribution.
We target various current systems, including the
recent Intel ``Nehalem'' and AMD ``Shanghai'' processors.

In recent years, the performance of various SpMVM algorithms has been
evaluated by several groups~\cite{GKAKK08,WOVSYD09,BG09p}.
Covering different matrix storage formats and implementations on
various types of hardware, they reviewed a more or less large number
of publicly available matrices and reported on the obtained
performance.
Their results account for the crucial influence of the sparsity pattern 
on the obtained performance:
Similar to a fingerprint, the positions of the non-zero entries in the matrix
determine which storage and multiplication scheme is best suited and how
fast this operation can be performed on a given platform.

We will take a different approach by isolating the individual
contributions to the run time of the SpMVM on a generic level, without 
resorting (at first) to specific matrices.
Based on the assumption that erratic, indirect memory access patterns
are decisive for SpMVM performance, an in-depth investigation 
of latency and bandwidth effects, hardware prefetching, and the influence of
non-local memory access on cache-coherent non uniform memory access (ccNUMA)
architectures (as implemented in almost all current HPC platforms) 
will clearly reveal the performance-limiting aspects on the multi-core
and multi-socket (node) level.
A successful performance model will be predictive for the expected performance
of various SpMVM implementations for a given matrix on the basis of its 
sparsity pattern, and give a hint to the respective optimal storage scheme.

This paper is organized as follows. In Section~\ref{sec:storage} we
describe the dominant sparse matrix storage schemes, CRS and JDS,
together with some promising refinements to
JDS\@. Section~\ref{sec:lowlevel} presents low-level benchmarks
for SpMVM-like access patterns and performance results for standard
and optimized storage layouts. In Section~\ref{sec:threaded} we
elaborate on multi-threaded SpMVM performance, both in multi-core
and ccNUMA contexts. Finally, Section~\ref{sec:conc} gives
a summary and outlook to future work.

\section{Common storage schemes}\label{sec:storage}

There is a variety of options for storing sparse matrices in memory~\cite{BDDRV00}.
While the memory requirements of the different formats are only marginally
different, there can be a substantial impact on SpMVM performance, as will
be shown below.
If there is prior knowledge about the sparsity pattern (symmetry features, 
dense subblocks, off-diagonal structures etc.), it can be exploited 
to generate a specialized format, suitable only for one type of physical
problem but optimal with respect to data access properties. In absence 
of such additional information one is limited to general 
storage schemes.
Among these, the most popular one is the compressed row storage (CRS)
format, which is also the basis of many sparse solver packages
like, e.g., SuperLU~\cite{DGL99} and PETSc~\cite{PETSc}.
In a nutshell, the CRS format stores the matrix in three (one-dimensional)
arrays. 
For each non-zero element, its value ({\tt val}) and column index ({\tt
col\_idx}) are stored.
The third array ({\tt row\_ptr}) holds the offset into \verb.val. where 
the entries belonging to a new matrix row start. If the number of matrix
rows is $N_\mathrm{r}$, the SpMVM kernel code
for CRS looks like this:
\begin{lstlisting}
  do i = 1, %$N_\mathrm{r}$%
    do j = row_ptr(i), row_ptr(i+1) - 1
      resvec(i) = resvec(i) + val(j) * %invec(col\_idx(j))%
    enddo
  enddo
\end{lstlisting}
The success of the CRS format lies in its simplicity and the high
computational performance on most cache-based
architectures, which results from the inner loop having the
characteristics of a sparse scalar product with an algorithmic balance
of 10~bytes/Flop~\cite{HW08}\@. On vector
systems, however, this format has the serious drawback of a short
inner loop if the number of non-zeros per row is not at least a couple
of hundreds.

To make SpMVM run well on vector processors, alternative formats have
been conceived, the most prominent being ``jagged diagonals storage'' 
(JDS)\@. In JDS, large vector lengths are given priority
over minimizing data transfer:
In a first step the rows and columns of the matrix are permuted such
that the number of elements per row decreases with increasing row index.
All further calculations are then performed in this permuted basis.
In each row the permuted non-zero elements are shifted to the left.
The resulting columns of decreasing length are called \emph{jagged diagonals}
and are stored consecutively in memory. There is one
array for non-zeros ({\tt val}) and one for column indices ({\tt col\_idx}),
just as with CRS\@. 
A third array (${\tt jd\_ptr}$) holds the starting offsets of the
jagged diagonals in {\tt val} and {\tt col\_idx}\@.
If $N_\mathrm{j}$ is the number of diagonals, the SpMVM kernel code 
for JDS looks like this:
\begin{lstlisting}
  do diag = 1, %$N_\mathrm{j}$%
    diagLen = jd_ptr(diag+1) - jd_ptr(diag)
    offset = jd_ptr(diag) - 1
    do i = 1, diagLen
      resvec(i) = resvec(i) + val(offset+i) * %invec(col\_idx(offset+i))%
    enddo
  enddo
\end{lstlisting}
Since the inner loop has the characteristics of a sparse vector
triad with an algorithmic balance of 18~bytes/Flop~\cite{Sc00}, its computational intensity is smaller than with CRS, but
due to the large loop length it is much better suited for vector processors and
similar machines. 

In view of the changing 
supercomputing landscape, where traditional cache-based and vector
architectures are being replaced by hybrid, hierarchical systems
comprising multi-core chips and accelerator hardware, it is worthwhile
pursuing both CRS and JDS as potentially promising approaches to
storing general sparse matrices.
In order to adapt the JDS format to the  bandwidth-starved situation
on current multi-core architectures, we have to 
balance the vectorization aspect against the potential of cache
and register reuse. 
Blocking (``NBJDS'') and outer loop unrolling (``NUJDS'') techniques seem most
promising in this respect, and can reduce the algorithmic balance of JDS considerably, 
so that it eventually becomes equal to CRS balance~\cite{HW08}. 
Standard blocking cuts all jagged diagonals into blocks of a given size.
Instead of updating the complete result vector for each jagged diagonal 
as a whole, only the elements of the current block are processed for
all jagged diagonals that have entries in this block, to the effect
that the corresponding part of the result vector remains in cache.
During a block update, accessing a new jagged diagonal
requires skipping entries in the {\tt val} and {\tt col\_idx} arrays.
In the ``RBJDS'' scheme we avoid this non-contiguous access by storing
all elements of a block consecutively.
In the outer-loop-unrolled ``NUJDS'' scheme each element of the result
vector is updated by two (or more) jagged diagonals simultaneously.
If the unrolling factor equals the number of jagged diagonals, this
variant is identical to the CRS scheme, aside from working in the
permuted basis.

\begin{figure}
  \centering\includegraphics*[width=0.98\linewidth]{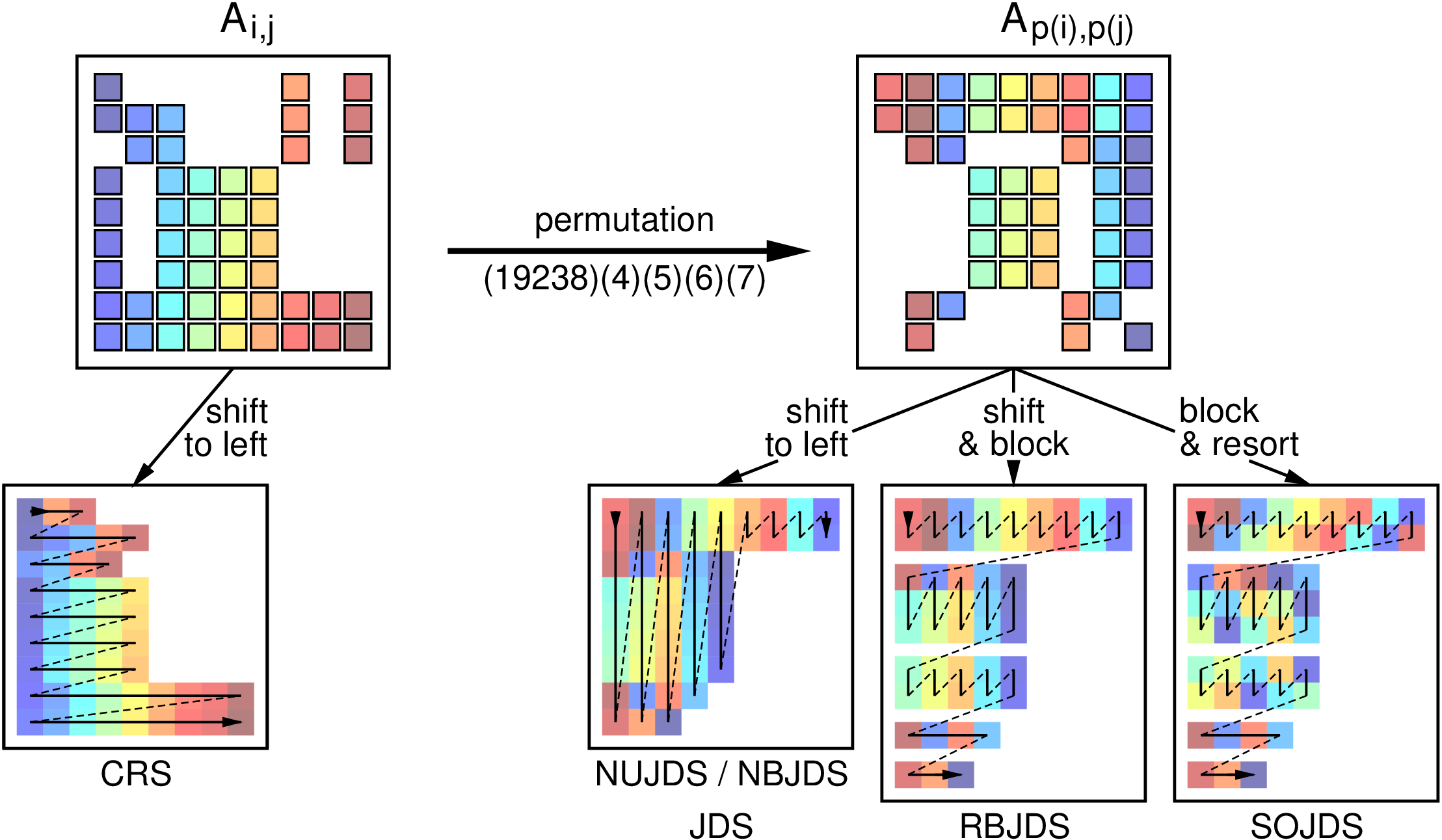}
  \caption{Construction of the {\tt val} and {\tt col\_idx} arrays for
           different storage schemes for a given matrix $A_{i,j}$.
           All JDS-flavored storage formats are based on the permuted
           matrix $A_{p(i),p(j)}$, with an additional sorting operation
           for SOJDS. The arrays within each sub-panel indicate 
           the storage ordering of the elements. While the storage
           pattern for JDS, NUJDS and NBJDS is identical, these methods
           differ in the corresponding access pattern}
\label{fig:storage_schemes}
\end{figure}

While all optimizations on the plain JDS format reduce memory traffic for 
writing the result, the discontinuous access to the input vector
remains a performance bottleneck.
Depending on the sparsity of the matrix and the anticipated number of required
SpMVMs it may be beneficial to use a more sophisticated initial ordering of the
elements.
In the ``SOJDS'' scheme the elements in a row are sorted such that within 
each column of a block the input vector is accessed with stride one (or as
close as possible)\@.
Figure~\ref{fig:storage_schemes} summarizes the ordering of the {\tt val} and 
{\tt col\_idx} arrays for the different storage schemes.

\section{Test bed}
\label{sec:testbed}
All benchmark tests were performed on three test systems:
\begin{description}
\item[\bfseries Woodcrest:] A two-socket UMA-type node based on 
	3\,GHz dual-core
	Intel Xeon 5160 processors (Core 2 architecture)
	with a 1333\,MHz frontside bus. The two cores in a 
	socket share a 4\,MB L2 cache.
	Measured STREAM Triad bandwidth is about 6.5\,GB/s\@.
\item[\bfseries Shanghai:] A two-socket ccNUMA-type node based on quad-core 
	AMD Opteron 2378 processors at 2.4\,GHz with two-channel DDR2-800
	memory. All four cores
	on a chip share a 6\,MB L3 cache.
	Measured STREAM Triad bandwidth is about 20\,GB/s\@.
\item[\bfseries Nehalem:] A two-socket ccNUMA-type node based on quad-core
	Intel ``Core i7'' (Nehalem) X5550 processors at 2.66\,GHz
	with 3-channel DDR3-1333 memory.
	The four cores on a chip share an 8\,MB L3 cache.
	Measured STREAM bandwidth is about 35\,GB/s\@.
\end{description}
For comparison we also obtained performance data on one node
of HLRB-II\@.

\section{Limitations of serial performance}\label{sec:lowlevel}

As compared to the peak performance of a processor, the performance for
a SpMVM is governed by memory access, and will usually be far less than
10\% of peak. 
To single out different aspects of this performance reduction, in a first step
we focus on basic operations that are the building blocks for the SpMVM in the
following sections.

\subsection{Basic sparse vector operations}

The inner loops of CRS and JDS differ in essence by the amount of data which is
written in each iteration.
While in the CRS case the result may be kept in a register and written to 
memory once after completing the inner loop, in the JDS case the whole
result vector is written to memory $N_j$ times.
From the point of view of data to be read, the two building blocks are
identical. 
Besides one (large) consecutive array, {\tt val}, the elements of the second
array, {\tt invec}, are determined indirectly by the indexing array {\tt
col\_idx}.
The access to {\tt invec} causes three different performance penalties,
which we will discuss by means of microbenchmarks.
First, the indirect addressing of {\tt invec} costs an additional 4 bytes per
iteration for reading the (dense) indexing vector.
Second, even if we replace the indirect addressing, {\tt invec(col\_idx(i))}, by
a direct access with non-unit stride, {\tt invec(k*i)}, the amount of transfered
data is significantly higher than in the dense case since for each entry 
a whole cache line is read.
Third, the indirect addressing may make efficient hardware prefetching almost
impossible. 
Therefore, the relevant restriction for the performance might be latency,
not memory bandwidth.

\begin{figure}
   \centering\includegraphics[width=0.47\linewidth,angle=-90,clip]{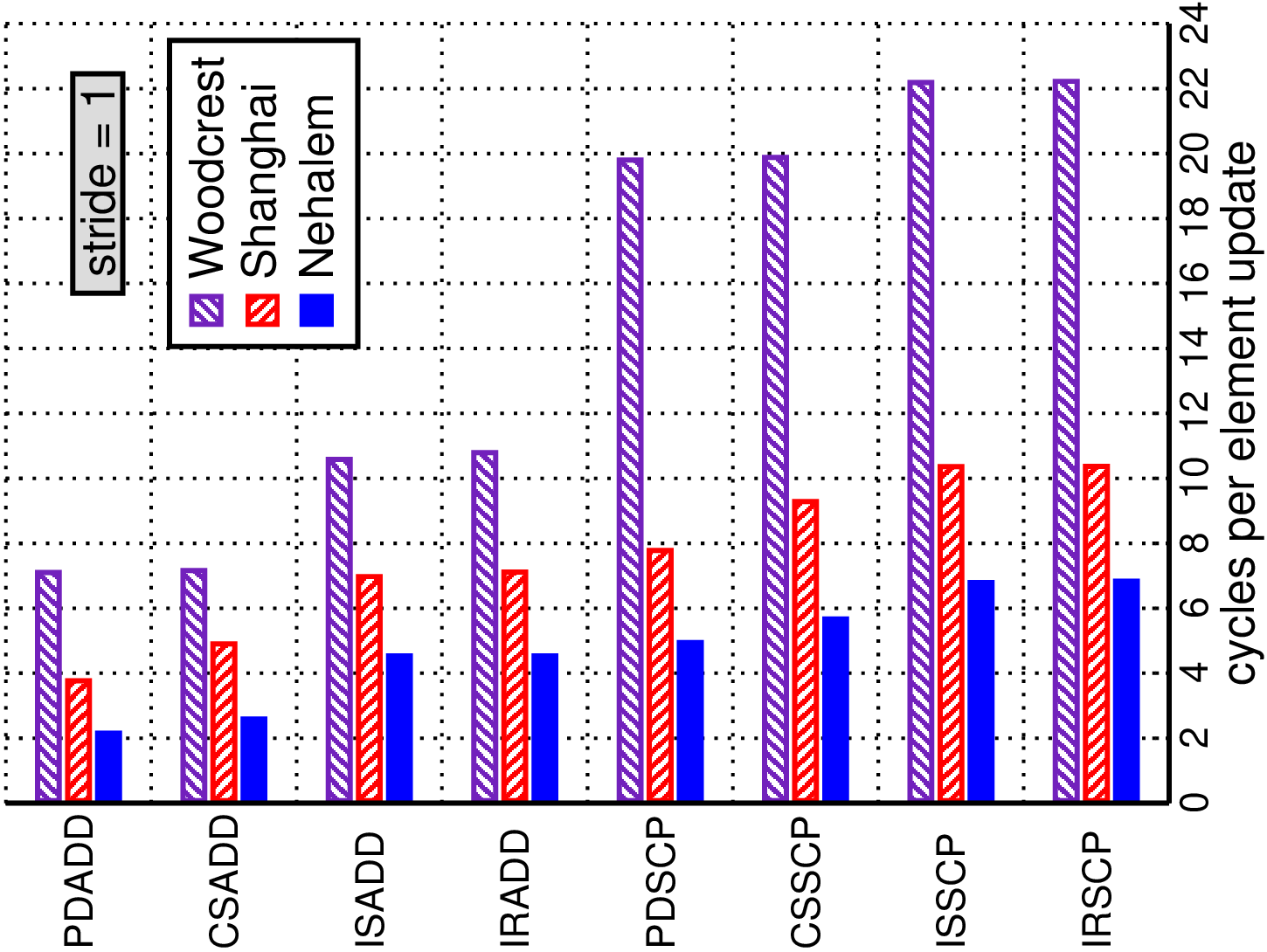}
   \centering\includegraphics[width=0.47\linewidth,angle=-90,clip]{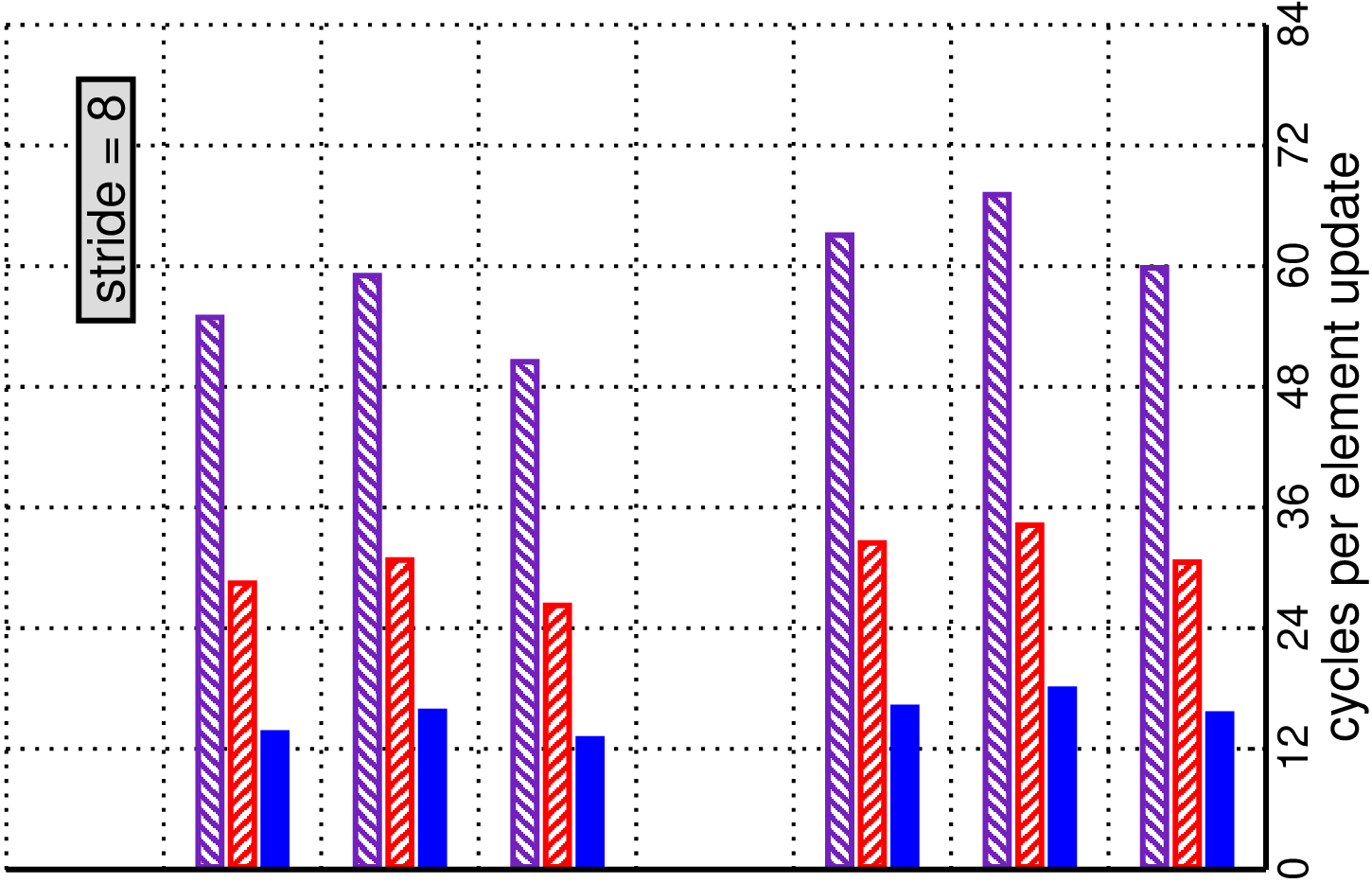}
   \centering\includegraphics[width=0.47\linewidth,angle=-90,clip]{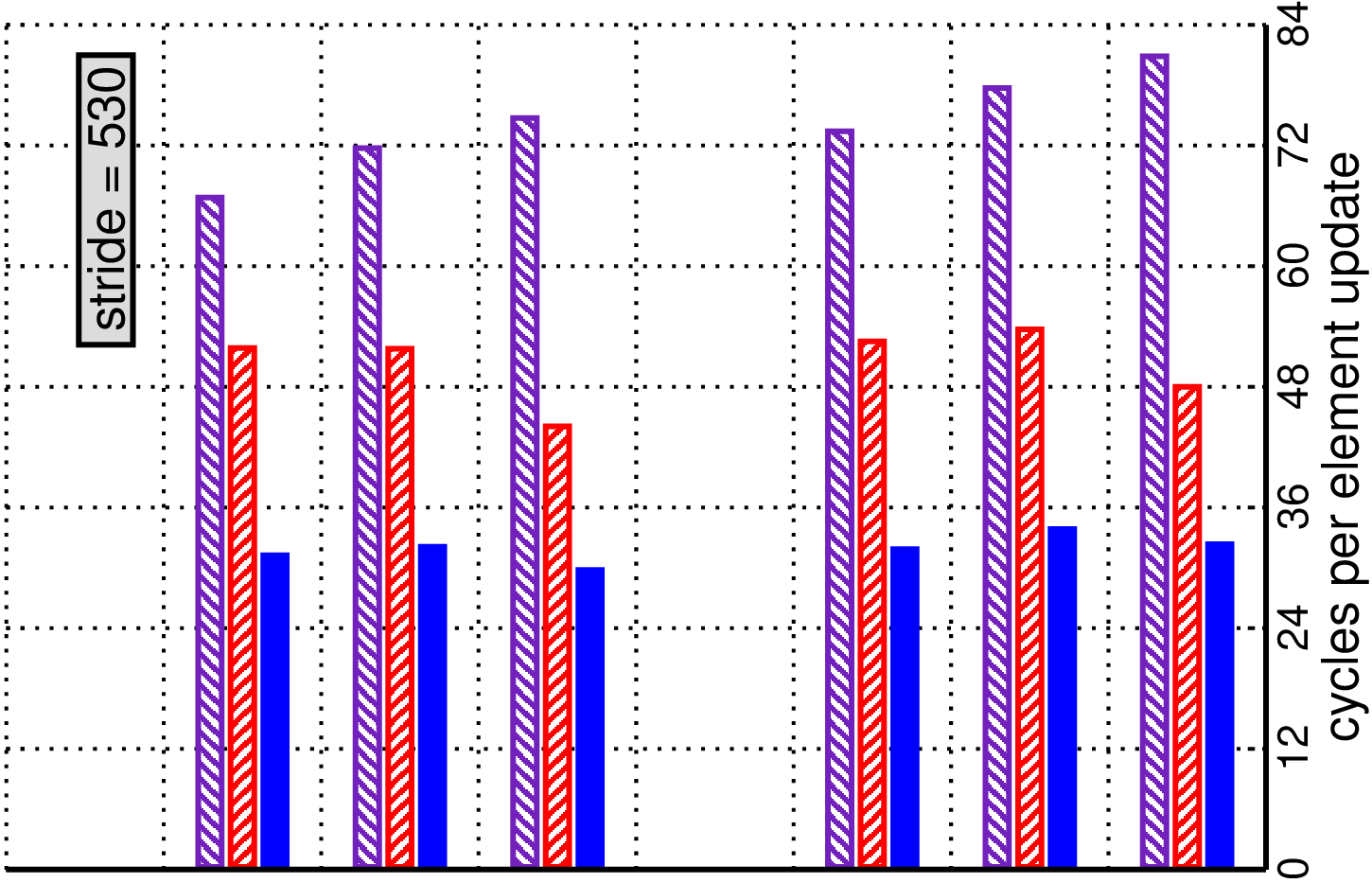}
   \caption{Performance of the basic sparse operations given in
            Tab.~\ref{tab:basic_operations} on different hardware architectures.
            Abstracting the used hardware the performance is given in required
            cycles per non-zero element update.  The considered strides
            correspond to dense packing, one entry per cache line and one entry
            per memory page. In the latter case we used stride $k=530$ in order
            to circumvent the performance penalties by cache trashing effects
            for $k=512$
  }
\label{fig:basic_operations}
\end{figure}

\begin{figure}
  \centering\includegraphics[width=0.49\linewidth,clip]{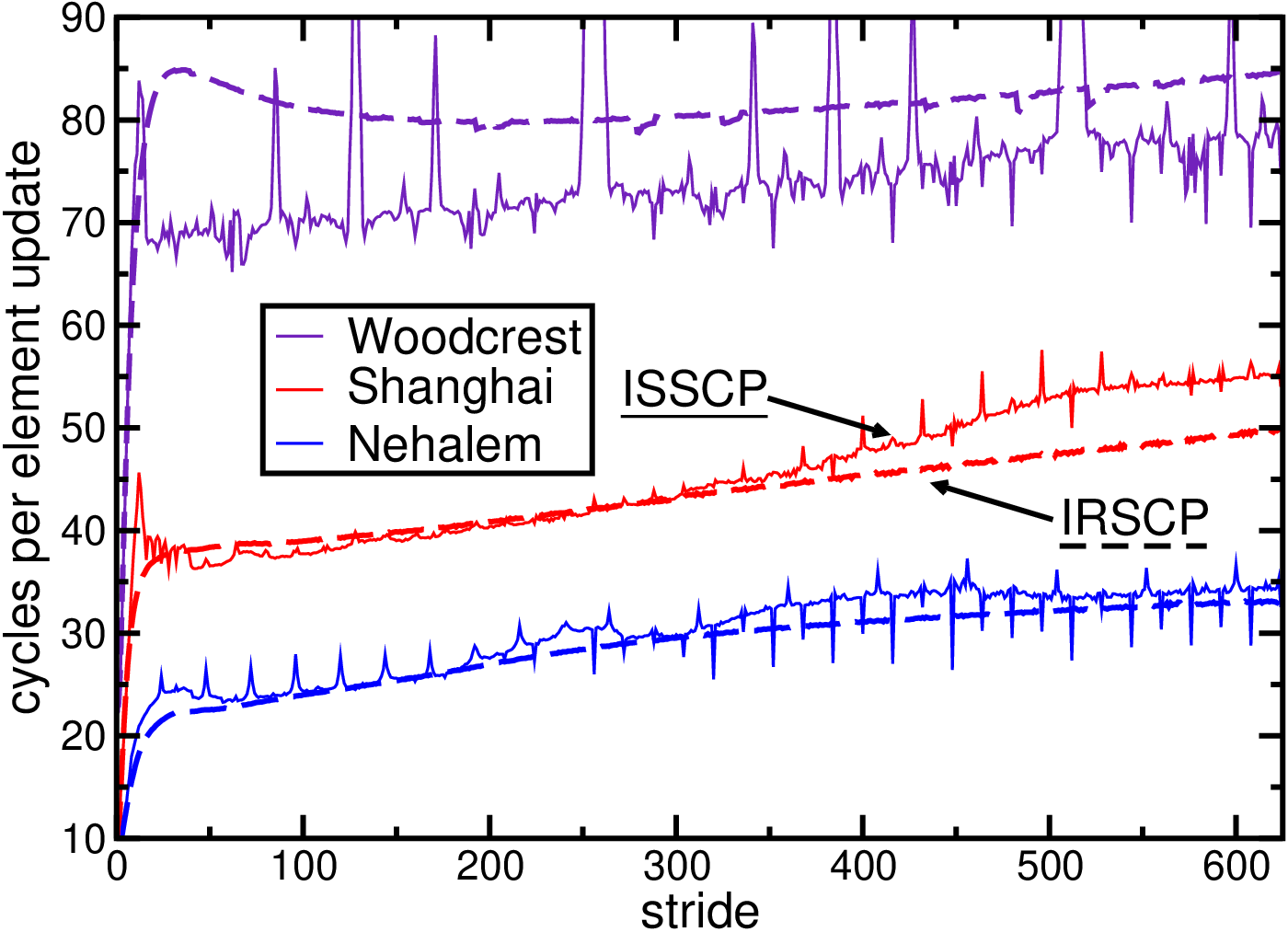}
  \centering\includegraphics[width=0.49\linewidth,clip]{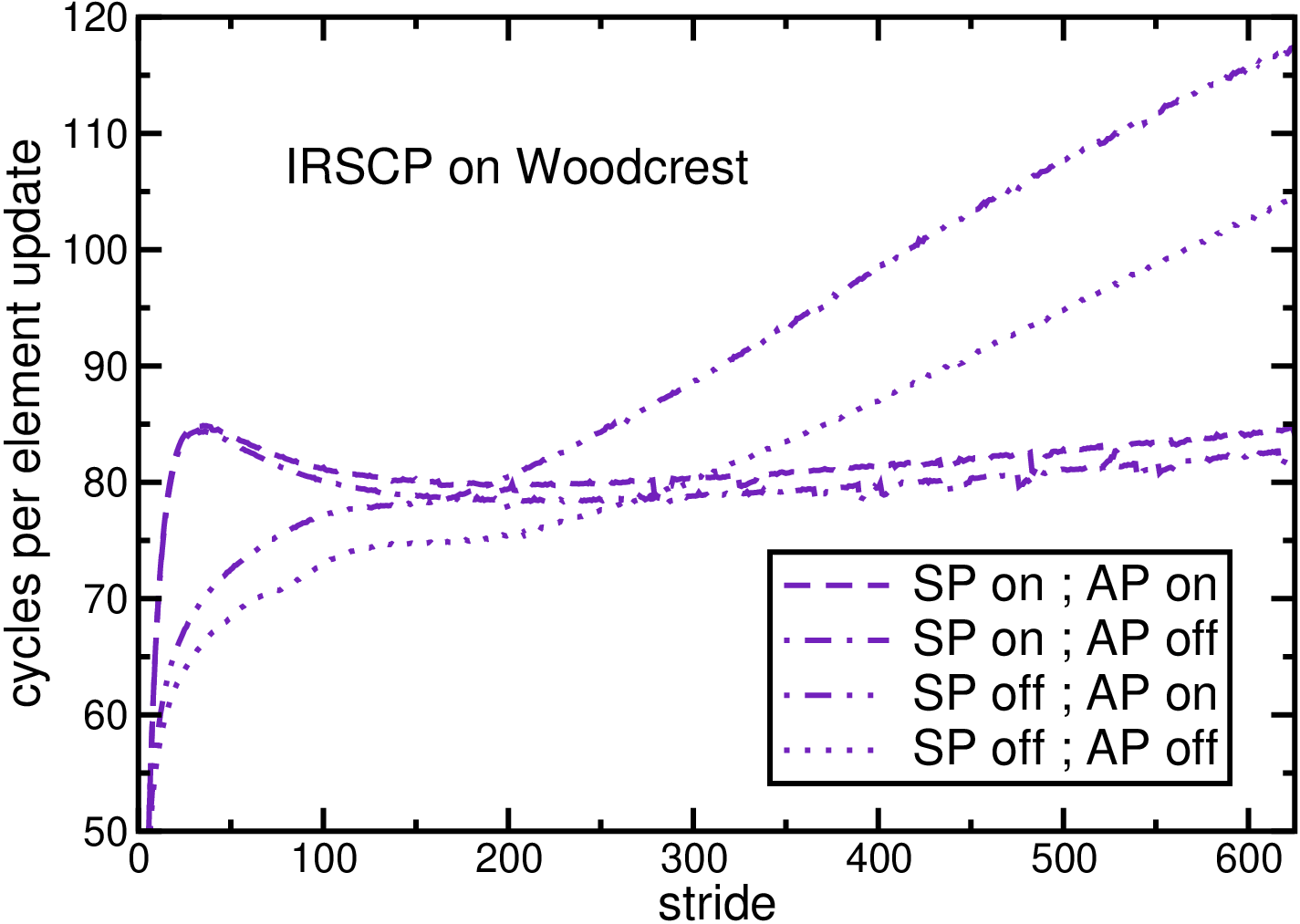}
  \caption{Left panel: Influence of the stride in the input vector on
           the performance of ISSCP (thin solid lines) and IRSCP (thick dashed
           lines). Right panel: Performance for deactivated strided
           (SP) and adjacent cache line (AP) prefetcher for IRSCP on Woodcrest}
\label{fig:stride_analysis}
\end{figure}

In an attempt to factor out the influence of loading the dense vector {\tt
val}, we abstract even more from the inner loop body by considering (indirect)
sparse vector additions and scalar products.
In Fig.~\ref{fig:basic_operations} we give the required cycles per element
update for the sparse scalar products and additions summarized in
Tab.~\ref{tab:basic_operations}.
Hereby the dense operations for stride $k=1$ serve as a baseline.
Note that the array lengths have been chosen such that the problem does not 
fit in any cache level.
Therefore the performance is limited by memory bandwidth, and the benefit
of using packed SSE loads ({\tt addpd}) instead of scalar ones ({\tt addsd}) in the
assembler code is marginal.%
\footnote{We do observe the expected factor of two for problem sizes which fit into
the cache.} 
The obtained performance using scalar loads is identical to the CSADD case
which means that the additional multiplication by the stride $k$ does not cause
a performance penalty as long as the innermost loop is sufficiently unrolled.
Indirect addressing causes an overhead of around 50\% for ISADD, which is in
accordance with the excess transferred data for the indexing array. 
Increasing the stride to $k=8$ we observe the anticipated performance
drop since for each element a whole cache line is read and seven eighths
of it are useless in this case.
This also holds for $k=530$, but the number of translation lookaside buffer 
(TLB) misses is drastically larger for this stride, leading to a measurable
penalty. 
A more detailed investigation of the performance of IRSCP and ISSCP 
on the stride is given in Fig.~\ref{fig:stride_analysis}(a). 
For the IRSCP benchmark we emulated these characteristics by generating a non-zero
element for each entry of {\tt invec} for which a drawn random number
is smaller than the threshold given by the inverse mean stride $p=1/k$. 

\begin{table}
   \centering
   \begin{tabular}{|c||l|l|}\hline
     & \centering{ADD} & SCP\\\hline\hline
         PD  & {\tt s = s + B(i)}      &  {\tt s = s + A(i) * B(i)} \\\hline
         CS  & {\tt s = s + B(k*i)}    &  {\tt s = s + A(i) * B(k*i)} \\\hline
     IS / IR & {\tt s = s + B(ind(i))} &  {\tt s = s + A(i) * B(ind(i))} \\\hline
   \end{tabular}
   \caption{Basic sparse vector operations, addition (ADD) and scalar product
            (SCP). Implementations are packed dense (PD), direct constant
            stride $k$ (CS) and indirect addressing (IS/IR). For the latter we
            distinguish two cases: either constant stride in the index array
            for IS, that is {\tt ind(i)=k*i} or random stride for IR. Here $k$
            is the mean stride.}
\label{tab:basic_operations}
\end{table}

Especially on the Woodcrest system, we encounter drastic performance drops for
ISSCP at strides which are multiples of powers of two.
On the other architectures these spikes also exist but are less severe.
They can be attributed to an effective reduction of cache size due to
cache trashing.
%
%
%
Randomizing the strides in IRSCP the distinct peaks disappear.
A remarkable feature that, however, survives is the bulge of reduced performance
for $k<25$, again most pronounced for the Woodcrest system.
To pin down its origins, we deactivated the hardware prefetching mechanisms of
the processor:
(i) The adjacent cacheline prefetch (AP), which loads two instead of one cacheline on
each miss, and (ii) the strided prefetcher (SP), which detects regular access patterns
and tries to maintain continuous data streams, hiding latency.
As shown in Fig.~\ref{fig:stride_analysis}, turning off the strided prefetcher,
the bulge vanishes and up to strides around $k\approx 200$ the performance
stays better than for active prefetchers.
However, this is not a general option since for larger strides SP is crucial
for the performance and disabling it results in massive penalties.
If we disable the adjacent cache line prefetcher, we observe 
an additional performance gain, either for activated or disabled SP,
the latter being more pronounced.
This can be attributed to the reduced memory traffic since now for each element
only one cache line is fetched from memory instead of two.
In general it is surprising how efficiently hardware prefetching works even with
irregular strides.

\begin{figure}
  \centering\includegraphics[width=0.98\linewidth,clip]{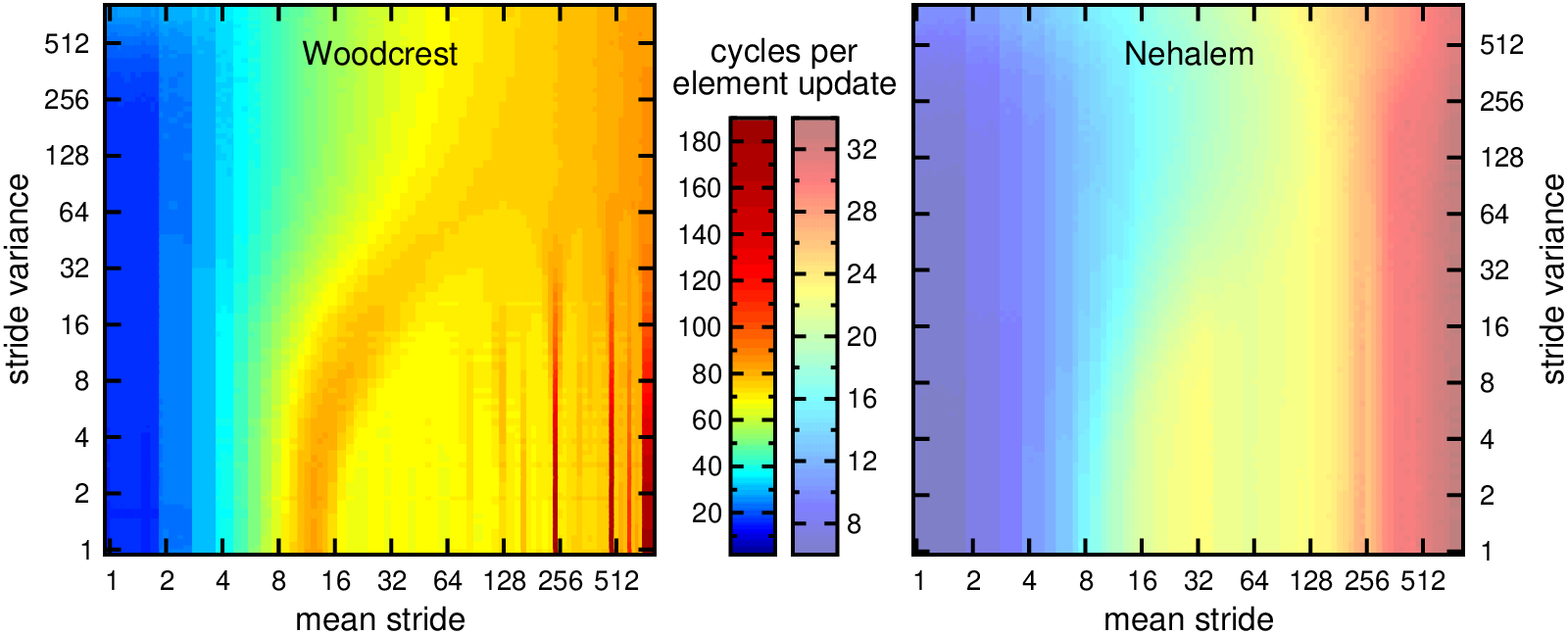}
  \caption{Performance of IRSCP for an {\tt ind} vector with strides drawn from
	   a Gaussian probability distribution with given mean value and
           variance of the stride}
\label{fig:stride_plane}
\end{figure}

Up to now we have focused on inner loop kernels, for which {\tt B}
is accessed in a strictly monotonic order with random positive strides only.
%
%
%
Clearly, the complete access pattern of the SpMVM kernels is not monotonic
since negative strides arise when the first element of a row refers to
an entry of {\tt invec} that has a lower index than the last entry of 
the previous row.
In order to investigate the influence of such backward jumps, we extend our IRSCP
studies to Gaussian-distributed strides (Fig.~\ref{fig:stride_plane}).
Fixing mean stride and variance of the distribution independently, we obtain a
refined control over distribution characteristics allowing for negative
strides provided the variance is large enough.
The obtained results for the Woodcrest system underline the findings shown in
Fig.~\ref{fig:stride_analysis}(a).
The peak structure seen in the ISSCP data is reproduced for small variances
while the stride jitter has only a minor effect on performance.
For large strides the smooth performance variations agree with the IRSCP data
in Fig.~\ref{fig:stride_analysis}(a) but the bulge observed there is missing
now.
We therefore attribute the bulge to the peculiarities of the chosen stride
distribution since its variance is not independent of the mean stride but grows
as $k(k-1)$.
Therefore the curves in Fig.~\ref{fig:stride_analysis}(a) correspond to a cut
along a tilted axis in Fig.~\ref{fig:stride_plane}, combining the properties for
small (large) variances at small (large) mean strides.
On the Nehalem system the fine performance structures are missing and we only
observe an overall decrease of the performance with increasing mean stride.

For Nehalem we also investigated the possibility of using non-temporal loads,
which bypass the cache hierarchy when moving data from memory to registers.
An instruction with the corresponding hint to the architecture ({\tt movntdqa})
was introduced in SSE4.1, which is supported by the Nehalem processor.
The non-temporal load may reduce the penalty connected with loading full 
cachelines of which only a fraction is actually used.
However we could not see any effect, positive or negative, perhaps due to the
hint-status of {\tt movntdqa}. 
On current processors it behaves like a standard load.

\subsection{Resulting performance for SpMVM}

\begin{figure}
  \centering\includegraphics[width=0.4\linewidth,clip]{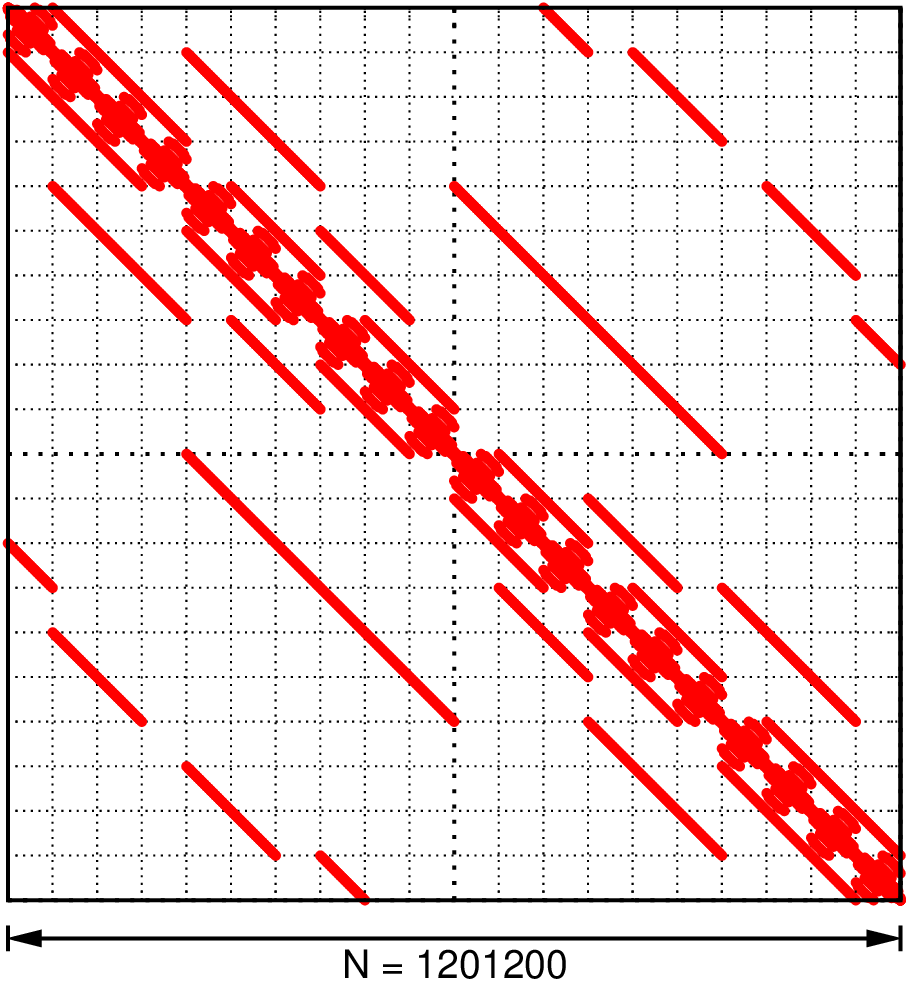}
  \centering\includegraphics[width=0.4\linewidth,clip]{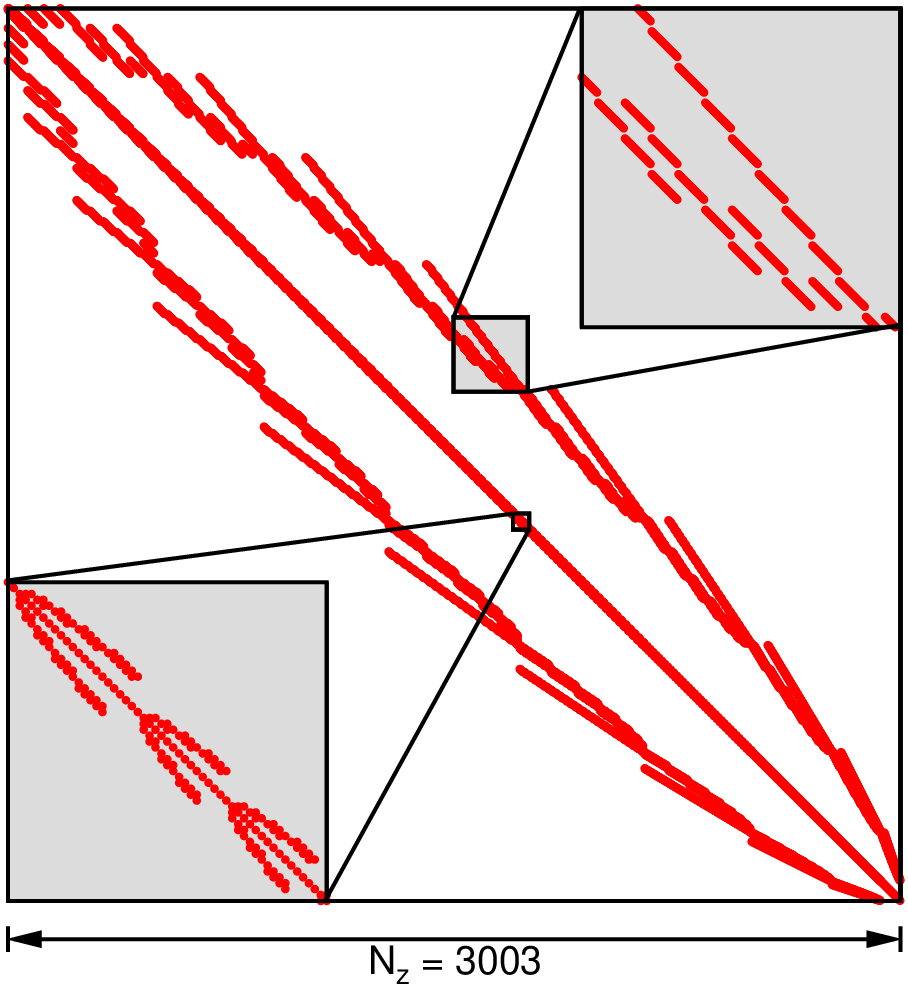}\\[2mm]
  \centering\includegraphics[width=0.8\linewidth,clip]{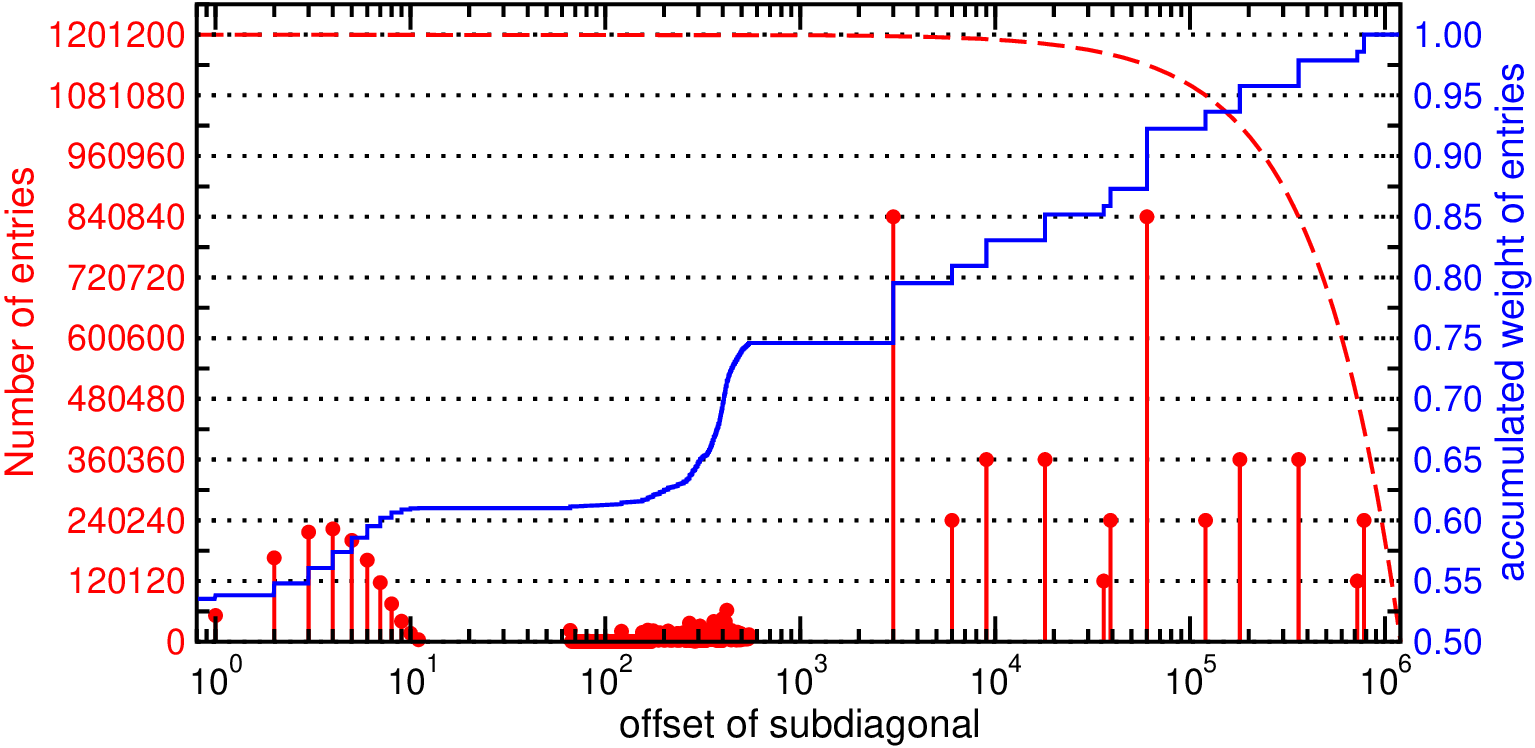}
  \caption{Top left panel: Sparsity pattern of the Holstein-Hubbard Hamiltonian
	   with dimension $N=1201200$.  Top right panel: detailed look at the matrix
	   structure near the diagonal. This panel magnifies the matrix by a
	   factor of $400$, i.e. its linear size corresponds to $1/20$ of a
	   dotted square in the left panel. In the further magnifying insets
	   ($N=54,250$) individual matrix entries can be distinguished.
	   Bottom panel: Compressed information on the matrix structure in
	   terms of offset and occupation of the subdiagonals. Due to
	   symmetry only the upper subdiagonals are shown. The dashed red
	   line gives the total (zero and non-zero) number of elements for each
           diagonal}
   \label{fig:sparsity_pattern}
\end{figure}
%
The low-level results from the previous section can be applied to a wide
variety of sparse numerical problems.
Since the sparsity pattern of a matrix influences the data access characteristics
in an essential way, this investigation will focus on a special physical problem, 
which is characterized by a Holstein-Hubbard Hamiltonian.
%
%
%
Two basic distributions of non-zero elements are present in the corresponding
sparse matrix:
From the sparsity pattern in Fig.~\ref{fig:sparsity_pattern} we see that
a considerable fraction of the matrix entries is concentrated in (rather dense) 
secondary diagonals. 
The remaining elements are scattered evenly over a band containing several hundred 
secondary diagonals, impeding the use of multi-diagonal storage schemes.
Such a split structure is typical for electron-phonon systems.
In addition, since the eigenvalues of a Hamiltonian are real-valued physical
observables the corresponding matrix is Hermitian (symmetric in the real case).
From the point of view of data transfer in the SpMVM this potentially allows for 
a further optimization which we do not investigate here.
Tracing back the matrix properties to the characteristics of our basic
benchmarks, the information contained in the sparsity pattern is too detailed.
In an attempt to compress this information we show in the lower panel of 
Fig.~\ref{fig:sparsity_pattern} the number of non-zero elements as a function
of their distance to the main diagonal together with the corresponding  
distribution function.
From the latter we see that about 60\% of the non-zero elements
are contained in the twelve outermost secondary diagonals. 
Each of those is a potential candidate for special treatment by a 
dense storage scheme.
Such hybrid implementations have been proposed in the literature~\cite{BG09p},
but we will not go into more detail on this aspect here.

\begin{figure}
  \centering\includegraphics[width=0.49\linewidth,clip]{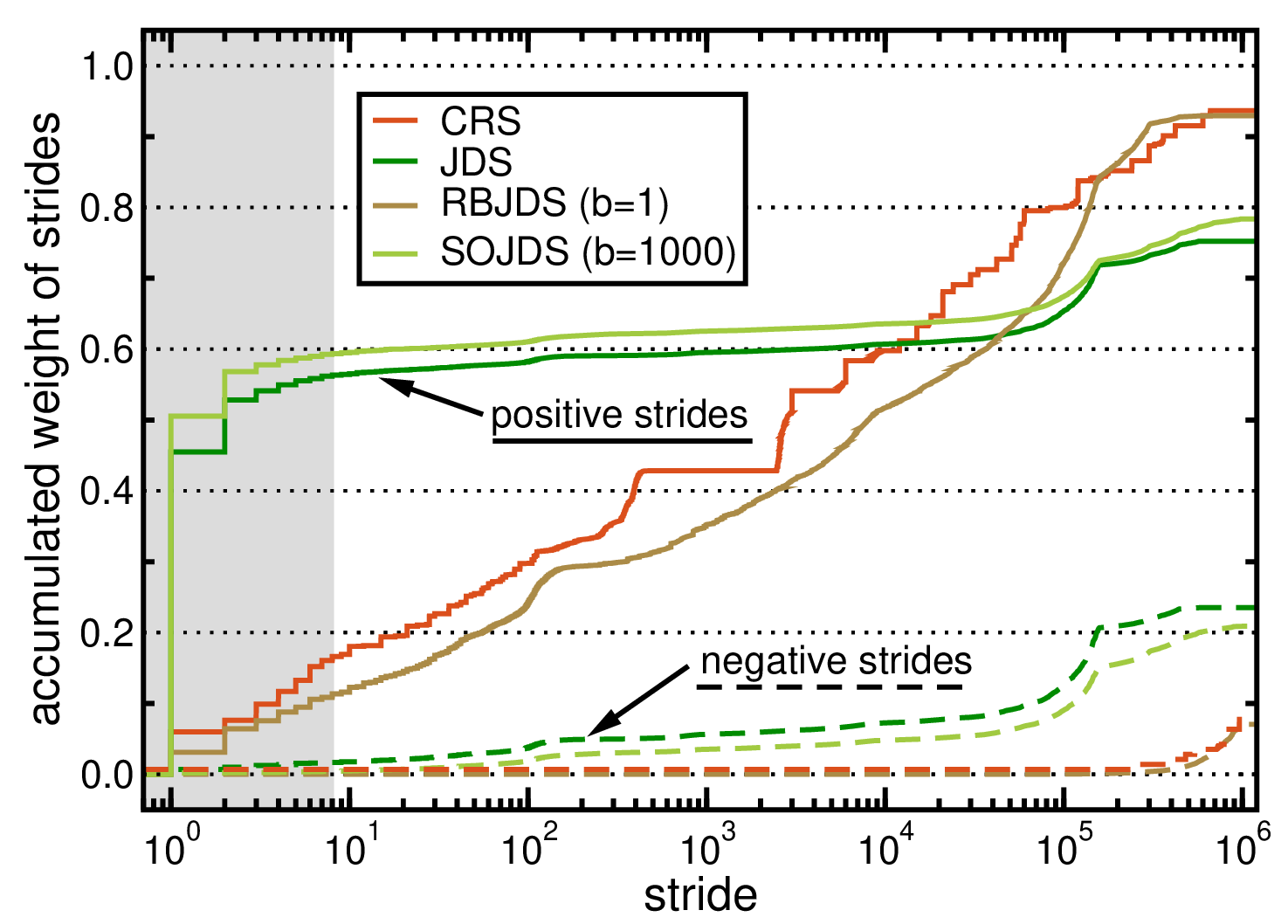}
  \centering\includegraphics[width=0.49\linewidth,clip]{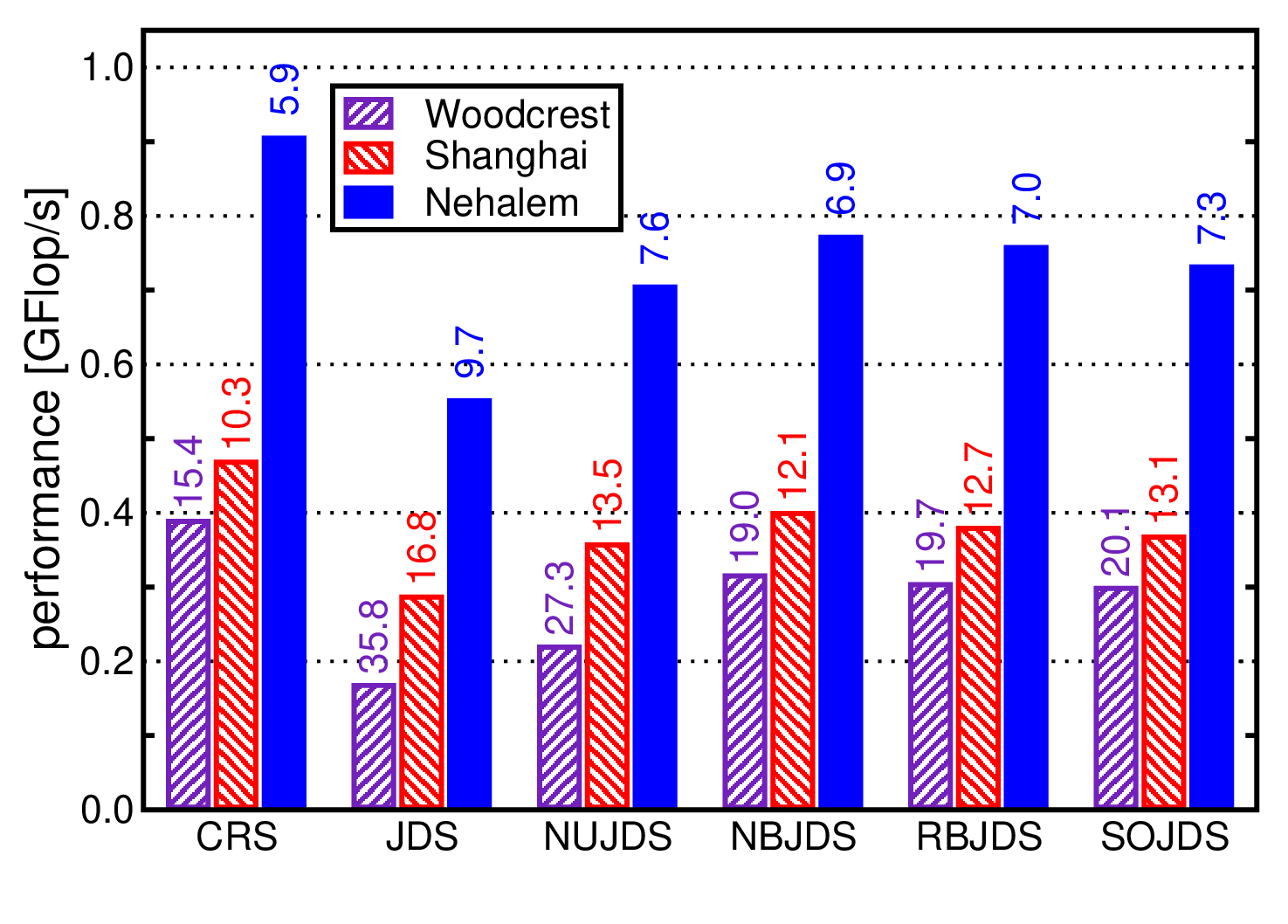}
  \caption{ Left panel: Distribution function of strides for the different
	    storage schemes.  The used block size for RBJDS (SOJDS) is 1
	    (1000). Solid (dashed) lines correspond to forward (backward) jumps
            in {\tt invec}.
	    Right panel: Serial performance for the complete SpMVM of the test
	    matrix using the different kernels. For the blocked versions the
	    block sizes have been chosen to optimize the performance (see
	    Fig.~\ref{fig:serial_performance_rrze1_blocksize}). In addition to the
	    number of floating point operations per second (Flop/s) also the
	    required cycles per element update are indicated for comparison to
	    previous results.}
   \label{fig:stride_dist_SpMVM}
\end{figure}
The stride distribution in the SpMVM kernel is influenced by more aspects than
just the secondary diagonals.
A further important factor is the storage scheme [see
Fig.~\ref{fig:stride_dist_SpMVM}(a)].
The stride distribution for CRS directly reflects the secondary diagonal
structure, with negative offsets only once at the beginning of a new row. 
Since the matrix has on average 14 non-zero elements per row, the accumulated
weight of backward jumps is around 7\%.
The positive strides are almost evenly distributed.
The required permutation of the matrix for setting up the JDS format
does not change this distribution significantly if the elements are
still accessed row by row.
This is the case for the RBJDS with block size 1.
Increasing the block size the distribution gradually approaches the limiting
case of JDS, which corresponds to a block size of the matrix dimension.
Here, the most pronounced feature is the increasing weight at small strides
such that almost 60\% of the strides are smaller than 64 bytes.
The drawback is a tripled amount of backward jumps.
SOJDS, which is optimized towards small strides, does not result in a
significant change in the stride distribution, even for large block size.
Comparing the resulting performance of all schemes
[Fig.~\ref{fig:stride_dist_SpMVM}(b)], the CRS still outperforms all 
JDS-flavored variants.
Interestingly, despite the improved memory access pattern, RBJDS and SOJDS
cannot outperform NBJDS if an optimal block size is chosen in all cases.
\begin{figure}
  \centering\includegraphics[width=0.98\linewidth,clip]{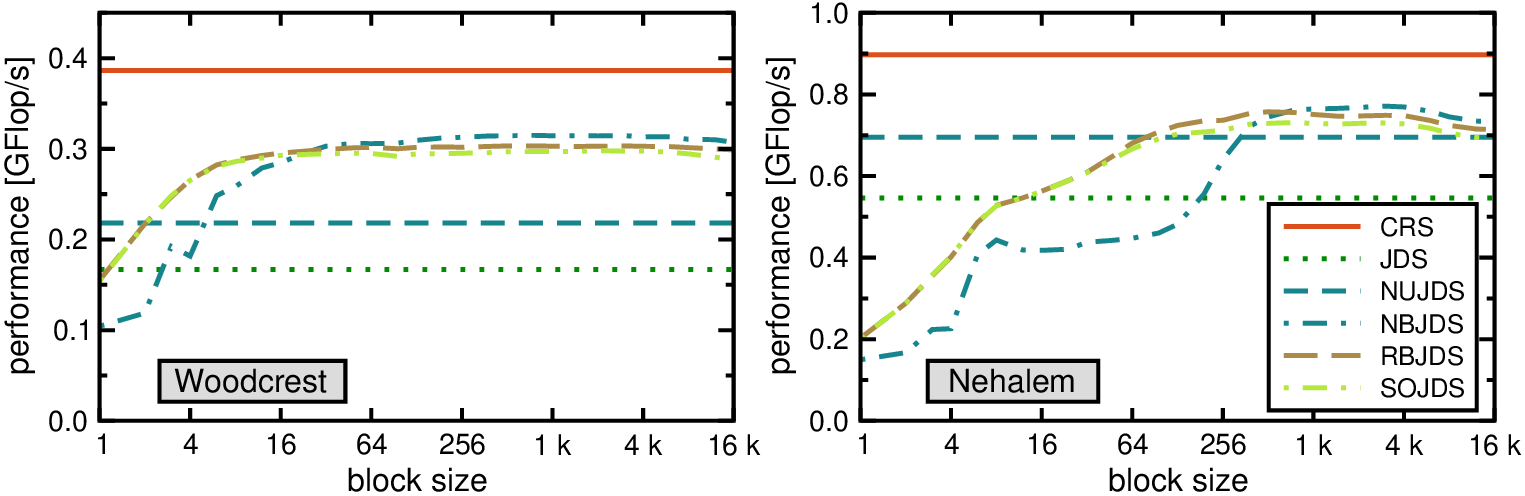}
  \caption{Block size dependence of the serial performance for the SpMVM with
	   the test matrix. For comparison also the performance for the schemes
           without blocking (CRS, JDS, NUJDS) is given. Data for Shanghai
           is not shown because the characteristics are very similar to 
           Nehalem, although on a lower level
           }
  \label{fig:serial_performance_rrze1_blocksize}
\end{figure}
As expected, the only benefit that can be drawn from the advanced blocked JDS
formats consists in a wider range of ideal block sizes (see
Fig.~\ref{fig:serial_performance_rrze1_blocksize}).
This may be important with parallel execution if load balancing becomes
an issue, which is however not the case for the considered test matrix.


\section{Shared-memory parallel SpMVM}\label{sec:threaded}

In order to fully exploit the potential of multiprocessor architectures,
parallel execution is mandatory.
In this work we consider OpenMP parallelization of our SpMVM code
on the systems described in Sec.~\ref{sec:testbed}.
Since all current commodity HPC systems are of the ccNUMA type, we 
distinguish clearly between intra-socket and inter-socket scaling 
behavior.
Of course, pinning all threads to the physical cores is crucial for obtaining
reliable performance data, because NUMA placement and bus contention effects
are important factors for the performance of all bandwidth bound algorithms. 
While for a serial process this can be easily accomplished using the {\tt taskset} 
command, pinning OpenMP threads is more involved. 
We implemented thread affinity by overloading the pthread library to ensure 
correct pinning of each thread created during program execution~\cite{MM_pin}.
%
%

\subsection{Intra-socket performance}

\begin{figure}
  \centering\includegraphics[width=0.98\linewidth,clip]{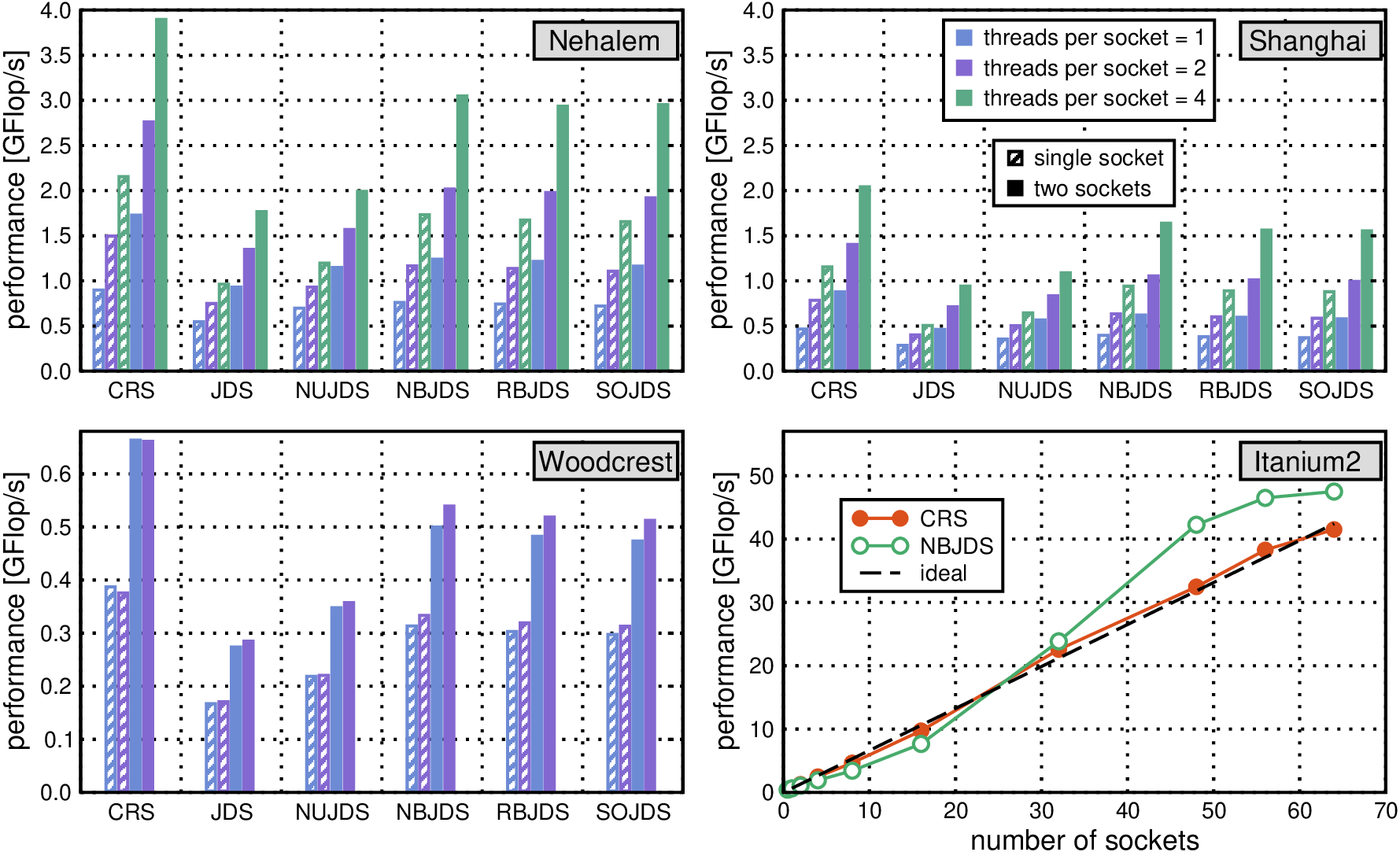}
  \caption{Performance of OpenMP parallel SpMVM kernels for one and two sockets
	   versus number of threads per socket. Data is based on static
	   scheduling and a block size of 1000. In the lower right panel the
	   measured (ideal) speedup on a HLRB II node is given by solid
           (dashed) lines. Two threads per node were used on this machine,
           and the CRS and NBJDS baselines were identical}
  \label{fig:performance_rrze1_OMP}
\end{figure}

In current multicore designs like Nehalem and Shanghai, a single thread is
not able to saturate the memory bandwidth of a socket.
The reasons for this are complex~\cite{THW09} and beyond the scope of this contribution.
For memory bound application like the SpMVM it is therefore obligatory to use
several threads per socket.
In order to minimize parallelization overhead, it is however desirable to use
the smallest number of threads that saturate the memory bandwidth.
In Fig.~\ref{fig:performance_rrze1_OMP} we compare the performance of the SpMVM
for our test matrix versus the number of OpenMP threads on different
architectures.
While for Nehalem and Shanghai the performance scales up to three threads per
socket, using a second thread per socket on Woodcrest results in no performance
gain.
As expected from the STREAM bandwidth numbers (see Sect.~\ref{sec:testbed}) Nehalem
outperforms Shanghai by a factor of roughly two.
The more recent AMD Istanbul processor has the same memory subsystem as Shanghai 
and is thus not expected to achieve better performance per socket.

\subsection{Inter-socket performance}
Using two sockets, performance on Woodcrest increases only by about 50\%, which 
is expected because this UMA-type FSB-based node architecture is known to
show scalability problems~\cite{HSZW09}.
The ccNUMA systems on the other hand scale much better, provided that proper
page placement is implemented by employing first touch initialization.
Apart from minor deficiencies which can be attributed to the access to {\tt invec}
our data reflects this feature (see Fig.~\ref{fig:performance_rrze1_OMP}).
Placement of the input vector is imperfect by design as non-local accesses from 
other NUMA domains cannot be avoided.
This is a property that strongly depends on the matrix structure, of course.

\begin{figure}
  \centering\includegraphics[width=\linewidth,clip]{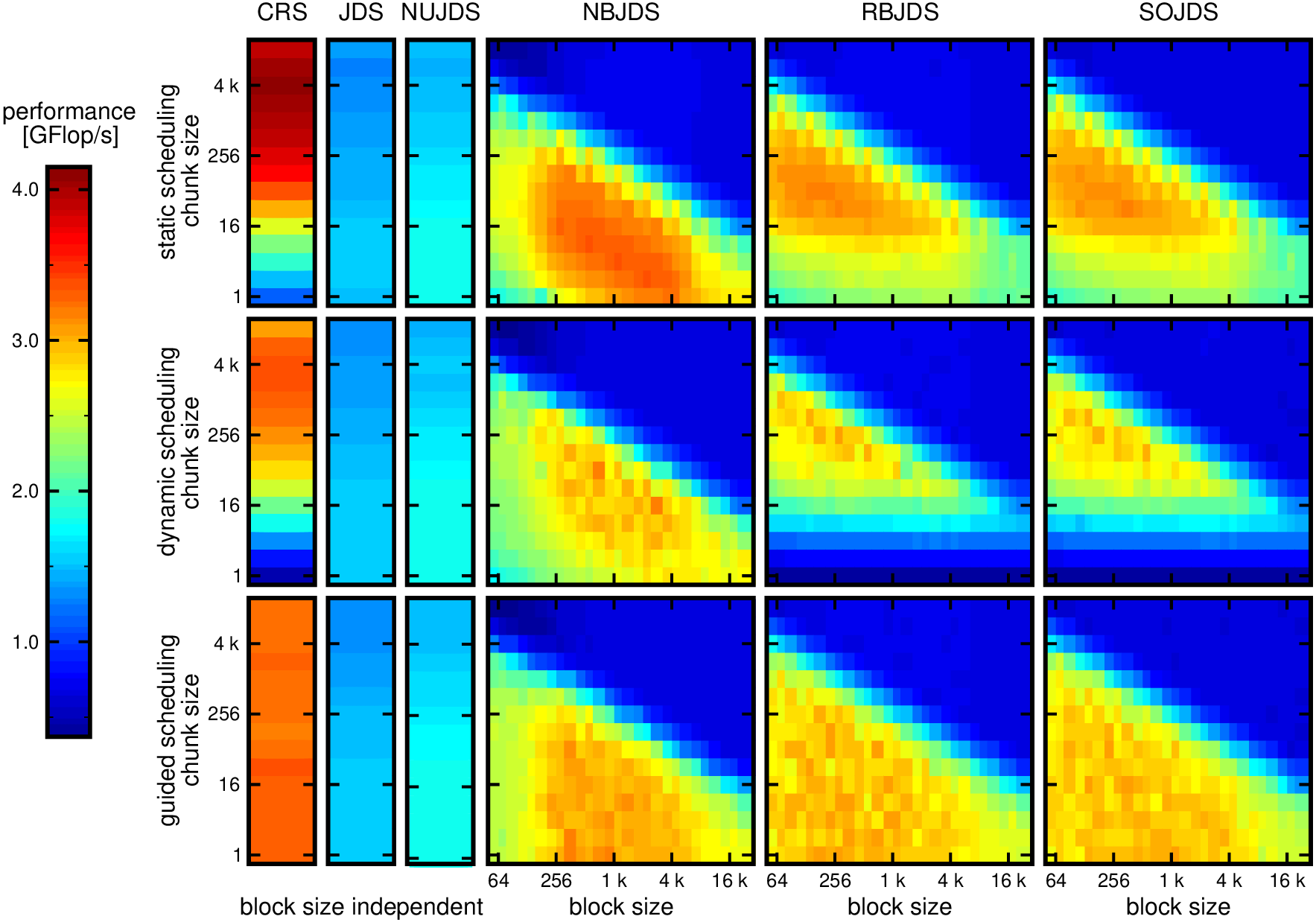}
  \caption{Performance of the SpMVM kernels as a function of block size, scheduling
           policy and scheduling chunk size for $2\times4$ threads on Nehalem.}
  \label{fig:performance_rrze1_OMP_schedule}
\end{figure}

Choosing the correct loop scheduling is vital on ccNUMA nodes.
Because of first touch placement, dynamic and guided scheduling are usually 
ruled out, except for strongly load imbalanced problems. 
In Fig.~\ref{fig:performance_rrze1_OMP_schedule} we investigate the impact 
of different OpenMP schedulings and chunk sizes using eight threads on the Nehalem 
system.

As expected, best overall performance is achieved with static scheduling
and the CRS format.
Small chunk sizes that lead to data blocks smaller than a memory page
are hazardous for performance since page placement becomes random. 
For all JDS flavors, using large block sizes together with large OpenMP chunks
leads to underutilized threads because the number of chunks becomes too small
(top right sector in all blocked JDS panels).

In summary, the superior performance of the CRS is maintained even for dual socket
ccNUMA systems.
For the considered Hamiltonian the possible benefits of load balancing 
by guided or dynamic loop scheduling does not outweigh the penalties 
from disrupted NUMA locality.

\subsection{HLRB-II scalability}

For comparison we performed scaling runs on a single node in the HLRB-II
bandwidth partition (two cores per locality domain).
In order to put the results shown in Fig.~\ref{fig:performance_rrze1_OMP} into
perspective, we must note that the whole matrix fits into the aggregated L3
caches of only 12 nodes.
Thus we expect two effects: (i) superlinear speedup might be observed and
(ii) JDS should have a significant performance advantage on many threads,
because of the negative performance impact of short loops (as in the 
CRS kernel) on the Itanium2.
Although CRS is slightly faster on 32 sockets or less, NBJDS dominates
for large thread counts, which confirms our predictions.

\section{Conclusion and outlook}
\label{sec:conc}

We analyzed the achievable performance of different SpMVM
implementations on current multicore multisocket architectures.
By a series of microbenchmarks, the basic operations of the SpMVM
have been investigated in detail.
The hardware prefetching mechanism of current x86 processors have
been shown to work unexpectedly well, even for moderately random
data access patterns.

Several SpMVM storage schemes have been benchmarked with respect to their
serial and OpenMP-parallel performance, using a Hamiltonian matrix from solid
state physics.
In summary, the CRS format outperforms the best cache-blocked JDS schemes
by at least 20\%.
Nevertheless we believe that JDS deserves further attention due to 
its long inner loop lengths, which might be advantageous on future 
processor and accelerator designs.

Future work will encompass a hardware counter analysis of SpMVM in 
order to get even more detailed information on its data access requirements.
In view of massively parallel systems distributed memory and hybrid
implementations will be thoroughly investigated.

\paragraph {\bf Acknowledgements} We thank J. Treibig and G. Wellein for
valuable discussion and acknowledge financial Support from KONWIHR II
(project HQS@HPC II)\@. 
We are also indebted to LRZ M{\"u}nchen for providing access to
HLRB-II\@.

\bibliography{ref}
\bibliographystyle{spphys}

\printindex

\end{document}